\documentclass[twocolumn]{aastex62}
\usepackage{float}	
\usepackage{hyperref}
\usepackage{amsmath}

\shorttitle{Calibrating the JWST filters as star formation rate indicators.}
\shortauthors{Senarath et al.}
\date{today}
\hypersetup{colorlinks=true, linkcolor=blue, citecolor=blue}

\begin{document}
\email{madhooshi.senarath@monash.edu}
\correspondingauthor{Madhooshi R Senarath}


\title{Calibrating the James Webb Space Telescope filters as star formation rate indicators.}


\author[0000-0003-1905-5426]{Madhooshi R. Senarath}
\affiliation{School of Physics and Astronomy, Monash University, Clayton, Victoria 3800, Australia}
\affiliation{Monash Centre for Astrophysics, Monash University, Clayton, Victoria, 3800, Australia}
\author[0000-0002-1207-9137]{Michael J. I. Brown}
\affiliation{School of Physics and Astronomy, Monash University, Clayton, Victoria 3800, Australia}
\affiliation{Monash Centre for Astrophysics, Monash University, Clayton, Victoria, 3800, Australia}
\author[0000-0002-9871-6490]{Michelle E. Cluver}
\affiliation{Centre for Astrophysics and Supercomputing, Swinburne University of Technology, Hawthorne, Victoria, 3122, Australia}
\affiliation{Department of Physics and Astronomy, University of the Western Cape, Robert Sobukwe Road, Bellville, 7535, South Africa}
\author[0000-0002-2733-4559]{John Moustakas}
\affiliation{Department of Physics and Astronomy, Siena College, 515 Loudon Road, Loudonville, NY 12211, USA}
\author[0000-0003-3498-2973]{Lee Armus}
\affiliation{IPAC, California Institute of Technology, Pasadena, CA 91125, USA}
\author[0000-0002-4939-734X]{Thomas H. Jarrett}
\affiliation{Department of Astronomy, University of Cape Town, Private Bag X3, Rondebosch, 7701, South Africa}





\begin{abstract}

We have calibrated the 6.5~m James Webb Space Telescope (JWST) mid-infrared filters as star formation rate indicators, using JWST photometry synthesized from \textit{Spitzer} spectra of 49 low redshift galaxies, which cover a wider luminosity range than most previous studies. We use Balmer decrement corrected $\rm{H\alpha}$ luminosity and synthesized mid-infrared photometry to empirically calibrate the \textit{Spitzer}, WISE and JWST filters as star formation rate indicators. Our \textit{Spitzer} and WISE calibrations are in good agreement with recent calibrations from the literature. While mid-infrared luminosity may be directly proportional to star formation rate for high luminosity galaxies, we find a power-law relationship between mid-infrared luminosity and star formation rate for low luminosity galaxies ($L_{\rm H\alpha} \leq 10^{43}~{\rm erg~s^{-1}}$). We find that for galaxies with a $\rm{H\alpha}$ luminosity of $\rm{10^{40}}~erg~s^{-1}$ (corresponding to a star formation rate of $\sim 0.055~{\rm M_\odot~yr^{-1}}$), the corresponding JWST mid-infrared $\nu L_{\nu}$ luminosity is between $\rm{10^{40.50}}$ and $\rm{10^{41.00}}~erg~s^{-1}$. Power-law fits of JWST luminosity as a function of $\rm{H\alpha}$ luminosity have indices between 1.17 and 1.32. We find that the scatter in the JWST filter calibrations decreases with increasing wavelength from 0.39 to 0.20~dex, although F1000W is an exception where the scatter is just 0.24~dex. 
\end{abstract}




\section{INTRODUCTION}

Galaxies grow via both star formation and galaxy mergers, so star formation rate (SFR) measurements are crucial for understanding galaxy evolution. SFR measurements can utilize spectra and imaging of both emission lines and continuum, and include (but are not limited to) line emission from $\rm{H\alpha}$ and $\rm{Pa\alpha}$ and continuum emission in the ultraviolet (UV), infrared (IR) and radio. These wavebands trace the presence or recent death of high mass stars with short lifetimes ($\lesssim 100~{ \rm{Myr}}$) \citep[e.g.,][ and references therein]{ken1998,Ken2012,Dav2015}. UV and $\rm{H\alpha}$ luminosities as a function of SFR can be predicted from theory \citep[e.g.,][and references therein]{ken1998}, but both can suffer greatly from dust attenuation, which can result in large uncertainties and systematics in measured SFRs. Mid-infrared (MIR) emission from dust and polycyclic aromatic hydrocarbons (PAHs) associated with star formation suffers little from dust obscuration, making it a potentially powerful SFR indicator. The main contaminant in MIR SFR calibrations is dust heated by old stellar populations that are unassociated with recent star formation, although the significance of this effect becomes less important for shorter MIR wavelengths \citep[e.g.,][]{Cal2010,Ken2012}. Critically, MIR SFR indicators require empirical calibration.


The 6.5~m James Webb Space Telescope (JWST), will be the most sensitive MIR telescope built to date, and will be able to measure SFRs for galaxies at $z < 3$. JWST's Mid-Infrared Instrument (MIRI) spans wavelengths between 5 - 30 $\rm{\mu m}$ using 9 filters with effective wavelengths of; 5.6 $\rm{\mu m}$, $7.7~{\rm \mu m}$, $10~{\rm \mu m}$, $11.3~{\rm \mu m}$, $12.8~ {\rm \mu m}$, $15~{\rm \mu m}$, $18~{\rm \mu m}$, $21~ {\rm \mu m}$ and $25.5~{\rm \mu m}$ \citep{Bou2015}. For point sources with exposure times of $\rm{10^4}~s$, JWST has a sensitivity of $\sim 10^{-6}~{\rm Jy}$ for $8-30~{\rm \mu m}$ (signal-to-noise ratio of 10), far deeper than \textit{Spitzer} and WISE, which have sensitivities $\sim 10^{-5}~{\rm Jy}$ and $\sim 10^{-3}~{\rm Jy}$ respectively \citep{Gla2015}.

In this letter we calibrate the JWST mid-infrared filters as SFR indicators. As JWST has yet to be launched, we calibrate the JWST mid-infrared filters using photometry synthesized from \textit{Spitzer} spectra of low redshift galaxies \citep{Bro2014,Bro2017}. To validate our approach, we also calibrate the \textit{Spitzer} and WISE mid-infrared filters using the same methods and spectra, enabling direct comparison with prior literature. \citet{Bat2015} have also calibrated six MIRI filters using a sample of high luminosity galaxies with SFRs of $\sim$ 1 - 10 $\rm{M_{\odot}}~{\rm yr^{-1}}$, whereas our calibrations are made using lower luminosity galaxies that cover a wider luminosity range.

Throughout this letter we use AB magnitudes, and a Hubble constant of $H_0 = 71.9~{\rm km~s^{-1}~Mpc^{-1}}$ \citep{Bon2017}. We also adopt a \citet{Kro2001}\footnote{In this paper we convert from $L_{H\alpha,corr}$ to SFR using:  $SFR(\rm{M_{\odot}~yr^{-1}}) = 5.5\times10^{-42}{\it L_{H\alpha}}(\rm{erg~s^{-1}})$ for a \citet{Kro2001} \citep{Ken09}. To convert from \citet{Sal1955} IMF and \citet{Cha2003} IMF to \citet{Kro2001} IMF, the multiplicative factor is 0.70 and 1.20 respectively \citep{Ken09, Jas2015}.} initial mass function (IMF) for this letter, where we use the \citet{Ken09} conversion from $L_{H\alpha,corr}$ to SFR for a \citet{Kro2001} IMF. Throughout this paper we use $\nu L_{\nu}$ luminosities in units of $\rm{erg~s^{-1}}$, which is consistent with several recent SFR calibration papers \citep[e.g.,][]{Ken09,Jarr2011,Bat2015,Bro2017,Clu2017}. 


\section{SAMPLE SELECTION AND EMISSION LINE MEASUREMENTS}

To calibrate MIR SFR indicators, we need a sample of galaxies with accurate MIR luminosities and accurate measurements of a SFR indicator that can be calibrated from theory (i.e., ${\rm H\alpha}$, UV and TIR). We use the \citet{Bro2014} galaxy spectra and accompanying optical emission line measurements \citep{Bro2017} to calibrate the JWST MIRI filters as SFR indicators. \citet{Bro2014} use matched aperture photometry to rescale the optical and mid-infrared spectroscopy, thus mitigating aperture bias. The median rescale factors for our sample of 49 galaxies for \textit{Spitzer} $8~{\rm \mu m}$, $12~{\rm \mu m}$ and $20~{\rm \mu m}$ filters are 1.50, 1.52 and 1.18 respectively, and at $8~{\rm \mu m}$ have a range of 0.88 - 4.70 \citep{Bro2014}. The sample galaxies are at redshifts of $z<0.05$ and have absolute magnitudes of $-14.7 \geq M_g \geq -23.2$. We refer the reader to \citet{Bro2014} for photometric data of the galaxies and to Appendix A of \citet{Bro2014} for IRS spectra of the galaxies.

The optical emission line fluxes are taken from \citet{Bro2017}, which utilize optical spectra first presented in \citet{Mou2006} and \citet{Mou2010}. Following \citet{Mou2010}, these revised emission line fluxes were determined using modified versions of pPXF \citep{Cap2004} and GANDALF\footnote{Gas and Absorption Line Fitting Algorithm \citep{Sar2006,Bar2006}.} \citep{Sar2006} to model the stellar continuum and emission lines respectively. The \citet{Bro2014} sample contains 129 galaxies, but limiting this sample to those galaxies with nebular emission lines that satisfy a signal-to-noise threshold of five for ${\rm H\alpha}$ and ${\rm H\beta}$ reduces the sample to 82 galaxies.

\begin{figure}[htb!]
\begin{center}
\includegraphics[width = \columnwidth]{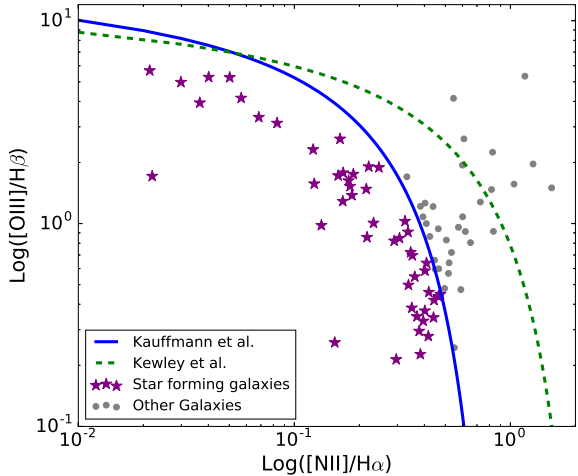}
\caption{BPT diagram of the 82 galaxies from \citet{Bro2014} where the measured ${\rm H\alpha}$ and ${\rm H\beta}$ emission line fluxes have a signal-to-noise greater than five. We plot both the \citet{Kew2001} and \citet{Kau2003} criteria, and use the \citet{Kau2003} criterion as it rejects more potential AGNs and LINERS. This leaves 49 star forming galaxies.}
\label{fig:BPTdiagram}
\end{center}
\end{figure}

To separate active galactic nuclei (AGNs) and low ionization nuclear emission-line regions (LINERS) from star forming galaxies, we use the Baldwin, Phillips and Terlevich \citep[][BPT]{Bal1981} diagnostic diagram. Figure~\ref{fig:BPTdiagram} shows the BPT diagram for our sample, along with the \citet{Kew2001} and \citet{Kau2003} criteria for selecting star forming galaxies. We have used the empirical \citet{Kau2003} criterion as it is more conservative, rejecting more galaxies that could potentially be AGNs or LINERs than \citet{Kew2001}. This reduces our sample size from 82 galaxies to 49 star forming galaxies.

Our SFR calibrations are anchored to ${\rm H\alpha}$ luminosities, which must be corrected for dust extinction. We used the extinction law of \citet{Car1989} for nebular emission and the \citet{Cal2000} starburst attenuation curve for the stellar continuum (although we don't use the dust corrected stellar continuum in this particular paper). We adopt the intrinsic emission line flux ratios of \citet{Ost1989} for Case B recombination with an effective temperature of $10,000$ K and electron density of $n_e$ = 10$^2$ cm$^{-3}$, giving an H$\alpha$ to H$\beta$ ratio of 2.86.





\section{JWST PHOTOMETRY}

As JWST has yet to be launched and take observations, our SFR calibrations rely on photometry we have synthesized from \textit{Spitzer} MIR spectroscopy of nearby galaxies \citep{Bro2014}. As \citet{Bro2014} includes MIR spectroscopy from \textit{Spitzer} and MIR photometry from \textit{Spitzer} and WISE, we can validate our approach by determining new SFR calibrations for \textit{Spitzer} and WISE  with synthesized photometry and directly comparing to the \citet{Bro2017} calibrations that used directly measured photometry. Galaxy luminosities are determined using (synthesized) apparent magnitudes and distances, with redshift independent distances being used when available \citep[all taken from the compilation of][]{Bro2017}. It should be noted that our sample has considerable overlap with that of \citet{Bro2017}, but does not include the \citet{Bro2017} galaxies that do not have \textit{Spitzer} MIR spectroscopy. 

Synthetic apparent magnitudes were determined using:
\begin{equation}
m = -2.5log\left[\left(\int R(\nu) \frac{f_\nu(\nu)}{h\nu} \mathrm{d} \nu \right) \times \left(\int R(\nu) \frac{g_\nu(\nu)}{h\nu}\mathrm{d} \nu \right )^{-1} \right]
\label{eq:magnitude}
\end{equation}
where $R(\nu)$ is the filter response function (electrons per incident photon), $h\nu$ is the energy of a photon with frequency $\nu$, $f_{\nu}$ is the galaxy SED and $g_{\nu}(\nu)$ is an AB magnitude zero source, with a flux density of 3631 Jy \citep[e.g.,][]{Hog2002}. 


For SFRs measured using photometry, it is common to use $f_{\nu}$ and $L_{\nu}$ defined using the apparent magnitude and the effective wavelength of the filter. This definition differs from Equation~\ref{eq:magnitude}, and is given by; 
\begin{equation}
    f_{\nu} = 3631~\rm{Jy} \times 10^{-0.4m}
    \label{eq:fluxdensity}
\end{equation}
where the $m$ is the apparent magnitude. For JWST we use the effective wavelengths defined by \citet{Bou2015} while for \textit{Spitzer} and WISE we use the effective wavelengths provided by \citet{Bro2017}.

\begin{figure}[htb!]
    \centering
    \includegraphics[width = \columnwidth]{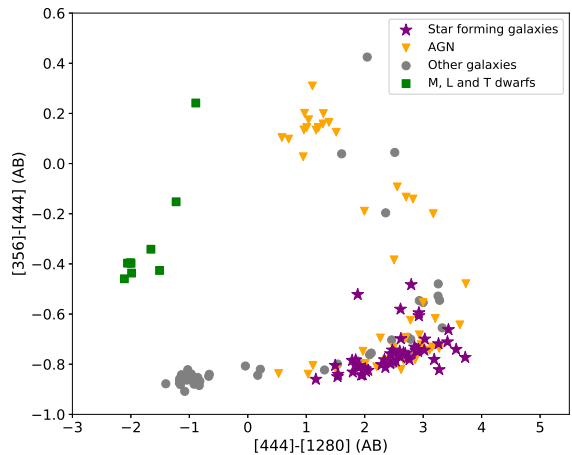}
    \caption{JWST NIRcam and MIRI color-color diagram. The AGNs in this figure are BPT selected AGNs from \citet{Bro2014} and quasars (Brown et al., in preparation), including AGNs dominated by host galaxy light that have mid-IR colors similar to star forming galaxies. Our diagram is similar to the WISE color-color diagram of \citet{Jarr2011}, as we have used filters with comparable effective wavelengths, and illustrates how MIR can separate powerful AGNs, passive galaxies, MLT dwarfs and star forming galaxies.}
    \label{fig:colour-colourNIRcam}
\end{figure}

To summarize the properties of our sample and for comparisons to prior literature, in Figure~\ref{fig:colour-colourNIRcam} we show a color-color diagram derived using MIRI and NIRcam filter curves  \citep{Gre2017}, which is similar to the WISE color-color diagram of \citet{Jarr2011}. We used all 129 galaxies from \citet{Bro2014} plus 16 additional AGN spectra from Brown et al. (in preparation).

New infrared SEDs for brown dwarfs and quasars were created by combining archival \textit{Akari} and \textit{Spitzer} spectra\footnote{http://cassis.sirtf.com/atlas/welcome.shtml and http://www.ir.isas.jaxa.jp}\textsuperscript{,}\footnote{The archival spectra used in Figure~\ref{fig:colour-colourNIRcam} are GJ1111, epsIndBa+Bb, J0036+1821, LHS3003, vB10, G196-3A, BRI0021-0214, J042348-0414, GJ1001A, 3C~273, 3C~351, Mrk~509, Mrk~876, PG~0052+251, PG~1211+143, PG~1415+451, PG~2349-014, Ton~951, 3C~120, Ark~120, Mrk~110, Mrk~279, Mrk~290, Mrk~590 and Mrk~817}. These were selected under the conditions that the spectra of a specific object appear in both archives and the {\it Spitzer} and {\it Akari} spectra had consistent flux densities at $\sim 5~{\rm \mu m}$. Unsurprisingly, the JWST color-color diagram closely resembles the WISE color-color diagram and illustrates the (well-known) utility of MIR photometry to select passive galaxies, star forming galaxies, brown dwarfs and quasars. A complete discussion of JWST color selection of different types of celestial object is beyond the scope of this letter, but will be expanded on in a future work.


\section{MIR FILTER CALIBRATIONS}


To validate the methods we use to calibrate the JWST MIRI filters, we first calibrate the WISE, \textit{Spitzer} IRAC and \textit{Spitzer} MIPS $\rm{24 \mu m}$ MIR filters using photometry synthesized from the \citet{Bro2014} spectra. These filters have been previously calibrated in the prior literature, which allows us to cross check our methodology. 

To calibrate a specific wavelength (or filter) as a SFR indicator, we model the relationship between the luminosity (at the relevant wavelength) and Balmer decrement corrected $L_{H\alpha}$ ($L_{H\alpha,corr}$). We used the least squares method to do this, which assumed a Gaussian scatter of the data about the line of best fit. The relationship between $\rm{H\alpha}$ luminosity and MIR luminosity is often modeled with a power-law \citep[e.g.,][and references therein]{Wu22005, Zhu2008, Ken09, Jarr2013, Lee2013,Bro2014,Clu2014, Bat2015, Bro2017,Clu2017}, and thus we also use this parameterization. 

\begin{figure*}[htb!] 
    \centering
    \includegraphics[width = \textwidth]{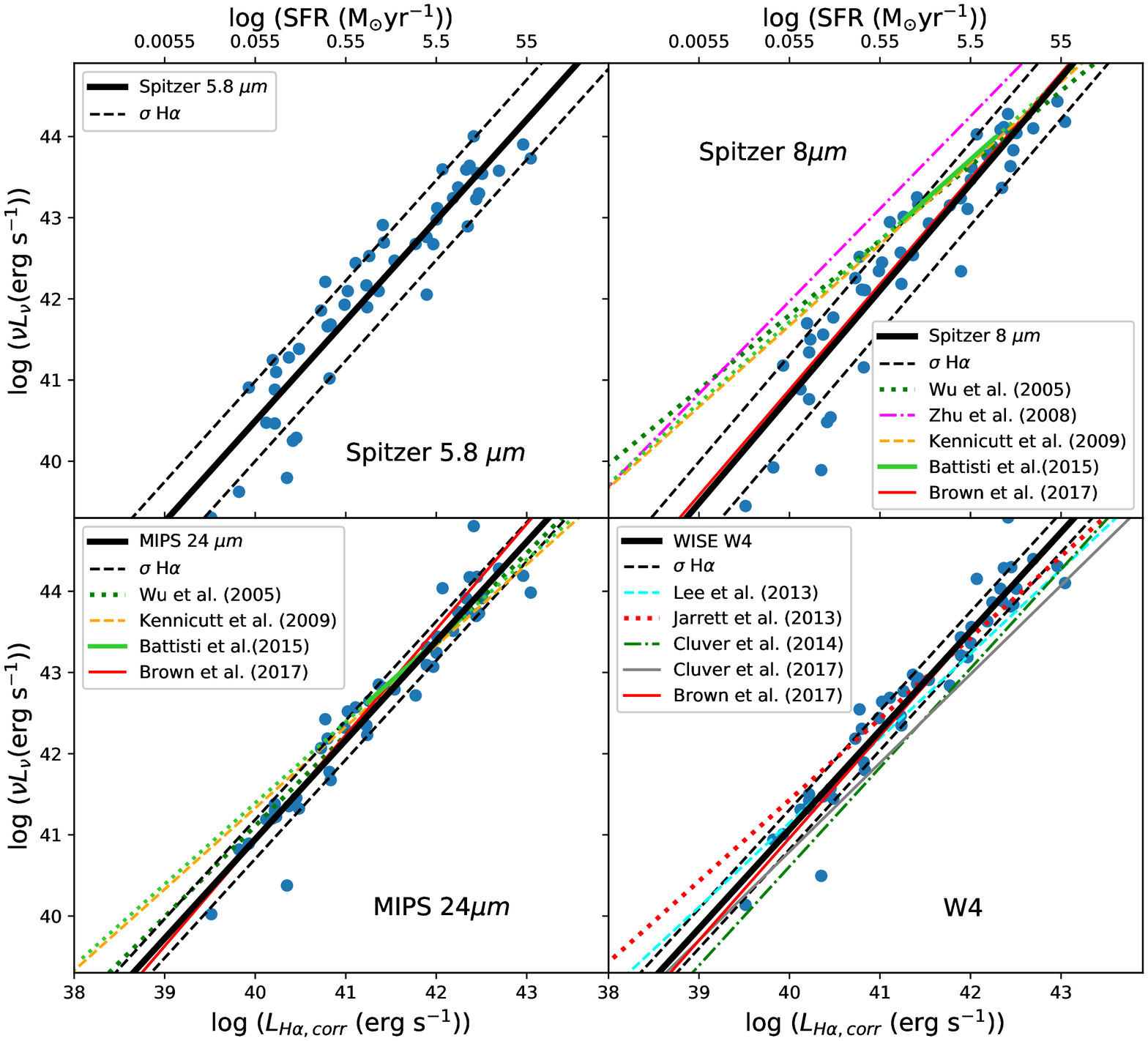} 
    \caption{Our \textit{Spitzer} and WISE SFR calibrations, along with calibrations from previous literature for comparison \citep{Wu2005,Zhu2008,Ken09,Lee2013,Jarr2013,Clu2014,Clu2017,Bat2015,Bro2017}. Our fits for \textit{Spitzer} and WISE and the equivalents from \citet{Bro2017} (shown in red) agree within each others' uncertainties, building confidence that our method can also be used for the JWST MIRI filters. The solid green line segments represent the MIRI calibrations of \citet{Bat2015} for the relevant SFR range, with the dotted green lines then extrapolating to lower SFRs.}
    \label{fig:irac}
\end{figure*}

As illustrated in Figure~\ref{fig:irac} and Table~\ref{table1}, for galaxies with a $\rm{H\alpha}$ luminosity of $\rm{10^{40}}~erg~s^{-1}$, we find the corresponding WISE and \textit{Spitzer} $\nu L_{\nu}$ is between $\rm{10^{40.49}}$ and $\rm{10^{41.37}}~erg~s^{-1}$. We find that the relationship between MIR luminosity and star formation rate is a power-law with indices in the range of 1.22 to 1.31. The WISE W4 normalizations and power-law indices agree to within 0.11 and 0.04~dex, respectively, of the \citet{Bro2017} estimates. Our \textit{Spitzer} normalizations and power-law indices agree to within 0.08 and 0.08~dex respectively of the \citet{Bro2017} estimates. These agreements build confidence in our methods, and their utilization for the JWST MIRI filters. \citet{Bro2017} include a number of dwarf galaxies in their sample that don't have MIR spectroscopy (and are thus excluded from our study), and we believe this is the main reason for the small systematic difference between our fits and those of \citet{Bro2017}, where we measure slightly smaller power-law indices. The dwarf galaxies extend to lower luminosities than covered by the sample we use for our calibrations.


The power-law fits for {\it Spitzer} $5.8 ~{\rm \mu m}$ and $8 ~{\rm \mu m}$ filters determined by other studies \citep[eg.,][]{Wu22005,Zhu2008} have smaller indices than those of \citet{Bro2017} fits and our fits. One reason for this discrepancy is we cover a broader range of $\rm{H\alpha}$ luminosities than much of prior literature, going down to $10^{39} ~{\rm erg~s^{-1}}$. Some studies also adopt a power-law index of 1, however dust content varies with galaxy mass \citep[e.g.,][]{Cal2000}, so this assumption is (at best) an approximation. For studies that use mostly high luminosity galaxies, the power-law indices are noticeably shallower, and create discrepancies when extrapolated to low luminosity dwarf galaxies \citep[e.g.,][]{Wu2005, Ken09}. 

We used $\rm{\chi^2}$ statistics to determine the uncertainties of our fits of luminosity vs $L_{H\alpha}$ relationship. The 1$\rm{\sigma}$ scatter about the line of best bit was determined by finding the range above and below fit that encompassed 68\% of the data. We present both the MIR scatter $\sigma_{\nu L\nu}$ and the $L_{H\alpha}$ scatter, $\sigma_{L_{H\alpha}}$, with the latter providing an estimate of the accuracy of MIR measurements of star formation. \citet{Bro2017} also measure the 1$\rm{\sigma}$ scatter about their calibration lines ($\sigma_{L_{H\alpha}}$ and $\sigma_{\nu L\nu}$) and we measure comparable scatter to \citet{Bro2017} for our calibrations. It should be noted that the scatter in the power-law fit is dominated by the true scatter of galaxies about the best-fit relationship, rather than being dominated by random errors (i.e., distance errors and photometric uncertainties) or systematic errors (i.e., Hubble constant errors, filter curve errors and zero point errors). 


\begin{figure*}[htb!] 
    \centering
    \includegraphics[width = 1\textwidth]{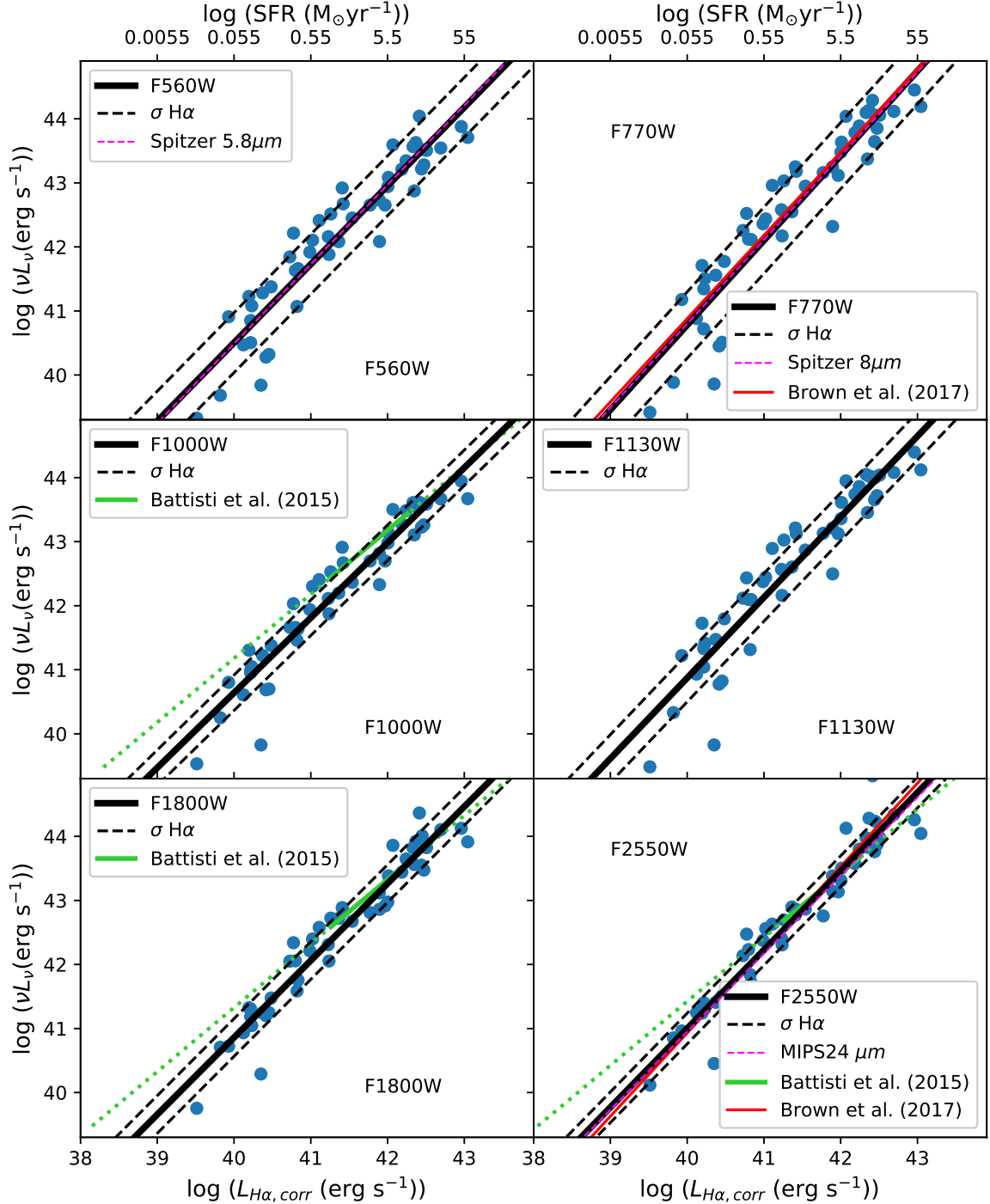}
    \caption{Calibrations of 6 out of the 9 JWST MIRI filters. The black dashed lines in the individual plots enclose 1$\rm{\sigma}$ of the SFR calibrators. The 1$\rm{\sigma}$ scatter decreases with increasing effective wavelength, dropping from from 0.39 dex to 0.20 dex. The solid green lines represent the MIRI calibrations of \citet{Bat2015}, where we have segmented the line so that it covers only the relevant SFR range, and extrapolated the line so that it covers the SFR of our galaxies (dotted green line).}
    \label{fig:JWSTcalibration}
\end{figure*}

Figure~\ref{fig:JWSTcalibration} shows our SFR calibrations for six of the nine JWST MIRI filters. In Figure~\ref{fig:JWSTcalibration}, we also plot our relations and those of \citet{Bro2017} for comparable {\it Spitzer} filters. We find good agreement for the F770W and F2550W filters with the normalizations and power-law indices agreeing within 0.05 dex and 0.01 respectively of our {\it Spitzer} calibrations. This builds confidence in our measurements for these wavelengths as well as the other 7 JWST MIRI filters. 

\begin{deluxetable*}{lcccc}
\centering
\tablecolumns{3}
\caption{Star formation rate indicator calibrations.}
\label{table1}
\tablehead{
  \colhead{Indicator} &
  \colhead{Fit to log($\nu L_{\nu}$)} &
  \colhead{$\sigma_{L_{H\alpha}}$} (dex)&
  \colhead{$\sigma_{\nu L_{\nu}}$ (dex)}&
  \colhead{Effective Wavelength}}
\startdata
\textit{Spitzer} $5.8~\mu m$ & (40.49 $\pm$ 0.10) + (1.24 $\pm$ 0.06) $\times$ $\left[{\rm log} (L_{H\alpha,corr}) - 40\right]$ & 0.40 &0.49& $5.73~\rm{\mu m}$ \\ 
\textit{Spitzer} $8~\mu m$ & (40.80 $\pm$ 0.11) + (1.31 $\pm$ 0.08) $\times$ $\left[{\rm log} (L_{H\alpha,corr}) - 40\right]$ & 0.39 &0.51& $7.87~\rm{\mu m}$\\ 
\textit{Spitzer} MIPS $24~\mu m$ & (40.95 $\pm$ 0.09) + (1.22 $\pm$ 0.07) $\times$ $\left[{\rm log} (L_{H\alpha,corr}) - 40\right]$ & 0.20 &0.24& $23.68~\rm{\mu m}$\\ \\ 
WISE W3 & (41.37 $\pm$ 0.13) + (1.24 $\pm$ 0.08)$\times$ $\left[{\rm log} (L_{H\alpha,corr}) - 40\right]$ & 0.31 &0.38& $11.56~\rm{\mu m}$\\ 
WISE W4 & (41.07 $\pm$ 0.07) + (1.22 $\pm$ 0.06)$\times$ $\left[{\rm log} (L_{H\alpha,corr}) - 40\right]$ & 0.20 &0.24& $22.8~\rm{\mu m}$\\ \\ 
JWST F560W & (40.50 $\pm$ 0.09) + (1.22 $\pm$ 0.08) $\times$ $\left[{\rm log} (L_{H\alpha,corr}) - 40\right]$ & 0.38 &0.47&$5.6~\rm{\mu m}$\\ 
JWST F770W & (40.78 $\pm$ 0.13) + (1.32 $\pm$ 0.07) $\times$ $\left[{\rm log} (L_{H\alpha,corr}) - 40\right]$ & 0.39 &0.52&$7.7~\rm{\mu m}$\\ 
JWST F1000W & (40.64 $\pm$ 0.10) + (1.17 $\pm$ 0.05) $\times$ $\left[{\rm log} (L_{H\alpha,corr}) - 40\right]$ & 0.24 &0.28&$10~\rm{\mu m}$\\ 
JWST F1130W & (40.87 $\pm$ 0.12) + (1.26 $\pm$ 0.08) $\times$ $\left[{\rm log} (L_{H\alpha,corr}) - 40\right]$ & 0.30 &0.37&$11.3~\rm{\mu m}$\\ 
JWST F1280W & (40.78 $\pm$ 0.08) + (1.24 $\pm$ 0.05) $\times$ $\left[{\rm log} (L_{H\alpha,corr}) - 40\right]$ & 0.30 &0.37&$12.8~\rm{\mu m}$\\ 
JWST F1500W & (40.74 $\pm$ 0.09) + (1.21 $\pm$ 0.05) $\times$ $\left[{\rm log} (L_{H\alpha,corr}) - 40\right]$ & 0.24 &0.29&$15~\rm{\mu m}$\\ 
JWST F1800W & (40.86 $\pm$ 0.09) + (1.20 $\pm$ 0.07) $\times$ $\left[{\rm log} (L_{H\alpha,corr}) - 40\right]$ & 0.24 &0.29&$18~\rm{\mu m}$\\ 
JWST F2100W & (40.92 $\pm$ 0.09) + (1.20 $\pm$ 0.07) $\times$ $\left[{\rm log} (L_{H\alpha,corr}) - 40\right]$ & 0.21 &0.25&$21~\rm{\mu m}$\\ 
JWST F2550W & (41.00 $\pm$ 0.10) + (1.23 $\pm$ 0.07) $\times$ $\left[{\rm log} (L_{H\alpha,corr}) - 40\right]$ & 0.20 &0.24&$25.5~\rm{\mu m}$\\ [1ex] 
\enddata
\end{deluxetable*}

For galaxies with a $\rm{H\alpha}$ luminosity of $\rm{10^{40}}~erg~s^{-1}$, the corresponding JWST MIRI $\nu L_{\nu}$ luminosity is between $\rm{10^{40.50}}$ and $\rm{10^{41.00}}~erg~s^{-1}$. Power-law fits of the data have indices between 1.17 and 1.32. Normalization and power-law fits, their uncertainties and scatter are all presented in Table~\ref{table1}. 

Using the less conservative \citet{Kew2001} BPT criteria, and after the removal of known AGN (such as IC~5298, NGC~1614, NGC~3079 and NGC~5033 etc.) and extreme outliers, we find normalizations and indices agree to within 0.13 dex and 0.02 of the power-law fits we previously determined using \citet{Kau2003} BPT selection criteria. We thus conclude our calibrations have a weak dependence on the chosen BPT criterion, with SFRs decreasing by $\sim$ 20\% (at fixed MIR luminosity) if we replace the \citet{Kau2003} criterion with the \citet{Kew2001} criterion.


Figure~\ref{fig:JWSTcalibration} and Table~\ref{table1} illustrate that the 1$\sigma$ scatter of the data about the power-law fits decreases with increasing wavelength from 0.38 dex for F560W to 0.20 dex for F2550W. This scatter is comparable to the scatter we measure for the \textit{Spitzer} and WISE filters as well as the scatter measured by \citet{Bro2017} and \citet{Clu2017} for filters of comparable wavelengths. The scatter is high in the shorter wavelength filters due to silicate absorption and peaks in PAH emission, including  strong features at $7.7~{\rm \mu m}$ and $\sim 12~{\rm \mu m}$ \citep[e.g.,][]{Bra2006, Clu2014}. Longer wavelengths are dominated by blackbody radiation from warm dust, resulting in less scatter.  

While the scatter generally deceases with increasing wavelength, the F1000W filter has a relatively low scatter of 0.24~dex, which is significantly less than filters with comparable effective wavelengths. This may be due to $10~ {\rm \mu m}$ being in the sweet spot where PAH emission is canceled out by silicate absorption. 

\citet{Bro2017} SFR calibrations included a number of dwarf galaxies with \textit{Spitzer} and WISE photometry but no \textit{Spitzer} IRS spectroscopy. That said, we observe scatter that is comparable to \citet{Bro2017} for the relevant wavelengths, even though we have used higher luminosity galaxies to calibrate the MIRI filters as SFR indicators. Our best fits have comparable normalization and power-law indices as the relevant \citet{Bro2017} fits (when applicable), so extrapolations of our relations should apply to dwarf galaxies. We conclude that the longer wavelength JWST MIRI filters, along with the F1000W filter, will provide the most accuracate SFR measurements.

In Figure~\ref{fig:JWSTcalibration} there are several low luminosity outliers (e.g., UGCA~166 and Mrk~475) and these are also seen in the \citet{Bro2017} calibrations. Lower luminosity galaxies used for the \citet{Bro2017} calibrations do fall along their power-law relations, so we don't believe the outliers in Figure~\ref{fig:JWSTcalibration} imply a breakdown of the overall power-law relation. As noted earlier, we were unable to include the lowest luminosity galaxies from \citet{Bro2017} in our calibrations as they lack {\it Spitzer} IRS spectroscopy. 

 
Our calibrations can be directly compared to those of \citet{Bat2015}, who presented calibrations of SFR indicators in $6 - 70~{\rm \mu m}$ wavelength range, including calibrations of six MIRI filters, some of which are plotted in  Figure~\ref{fig:JWSTcalibration}. To compare our calibrations with those of \citet{Bat2015} we use their conversion factors that correspond to $z=0$. Our sample includes relatively low luminosity galaxies whereas the \citet{Bat2015} sample is focused on galaxies with SFRs of $\sim 1 - 10~\rm{M_{\odot}}~{\rm yr^{-1}}$ with a few galaxies with SFRs $> 10~\rm{M_{\odot}}~{\rm yr^{-1}}$. \citet{Bat2015} have assumed that SFR is directly proportional to MIR luminosity, with a conversion factor that is determined via fits of synthetic observations of a redshifted composite spectrum. While this assumption may be correct for galaxies with high luminosity, we find that the relationship between $\rm{H \alpha}$ luminosity and star formation rate is a power-law. This can be clearly seen in Figure~\ref{fig:JWSTcalibration} where our relationship and that of \citet{Bat2015} intercept at higher luminosities, while at lower $\rm{H \alpha}$ luminosities the relationships from \citet{Bat2015} are ${\rm \approx 1~dex}$ higher in MIR luminosity than our calibrations. From this we conclude that the calibrations from \citet{Bat2015} will provide an accurate SFR measurement for higher luminosity galaxies, however our calibrations are more accurate for lower luminosity galaxies (i.e., dwarf galaxies). An obvious extension to our work will be to calibrate the JWST MIRI filters for a wider luminosity range, including high luminosity galaxies that fall in the range where prior calibrations have adopted a one-to-one relationship between MIR luminosity and SFR.

\section{SUMMARY}

We have calibrated the JWST, {\it Spitzer} and WISE MIR filters as SFR indicators using photometry synthesized from \textit{Spitzer} spectra and Balmer decrement corrected $\rm{H\alpha}$ luminosities, measured with scanned long-slit spectroscopy \citep{Mou2006,Mou2010,Bro2014,Bro2017}. Our galaxy sample covers a wide range of luminosities and thus our calibrations extend to lower luminosities than other studies such as \citet{Bat2015}. 
We verified our approach by comparing our {\it Spitzer} and WISE calibrations with those from the literature.
%
%

For {\it Spitzer}, WISE and JWST we find galaxies with an ${\rm H\alpha}$ luminosity of $10^{40}~{\rm erg~s^{-1}}$, the corresponding MIR $\nu L_{\nu}$ is between $10^{40.49}$ and $10^{41.37}~{\rm erg~s^{-1}}$. We find that the relationship between luminosity in MIR filters is approximated by power-laws with indices between 1.17 and 1.32. For the MIRI filters, the 1$\sigma$ scatter of the data about our power-law fits is between 0.39 and 0.20 dex, which are comparable to the scatter measured by \citet{Bro2017} for equivalent wavelength filters. This scatter decreases with increasing wavelength as PAHs emission and silicate absorption dominate at shorter wavelengths, whereas longer wavelengths have are dominated by blackbody radiation from dust. The exception to this is the $10~{\rm \mu m}$ filter (F1000W) which has a scatter of just 0.24 dex, which is significantly less than other JWST filters with comparable effective wavelengths.

\acknowledgments
This work is based in part on archival data obtained with the \textit{Spitzer Space Telescope}, which is operated by the Jet Propulsion Laboratory, California Institute of Technology under a contract with National Aeronautics and Space Administration. Support for this work was provided by an award issued by JPL/Caltech. This publication makes use of data products from the Wide-field Infrared Survey Explorer, which is a joint project of the University of California, Los Angeles, and the Jet Propulsion Laboratory/California Institute of Technology, funded by the National Aeronautics and Space Administration. This research is based on observations with AKARI, a JAXA project with the participation of ESA.


Facilities: \facility{Bok}, \facility{JWST}, \facility{\textit{Spitzer}}, \facility{WISE},\facility{\textit{Akari}}





\clearpage
\bibliographystyle{apj}
\bibliography{references}

\end{document}